\documentclass[12pt,preprint]{aastex}
\usepackage{amssymb}
\slugcomment{to appear in ApJS}
\shorttitle{VSOP Survey II}
\shortauthors{Lovell et al.}

\begin{document}

\title{The VSOP 5 GHz AGN Survey
    II. Data Calibration and Imaging}

\author{J.~E.~J.~Lovell           \altaffilmark{1},
        G.~A.~Moellenbrock        \altaffilmark{2},
        S.~Horiuchi               \altaffilmark{3}, 
        E.~B.~Fomalont            \altaffilmark{4},
        W.~K.~Scott               \altaffilmark{5},
        H.~Hirabayashi            \altaffilmark{6},
        R.~G.~Dodson              \altaffilmark{6},
        S.~M.~Dougherty           \altaffilmark{7},
        P.~G.~Edwards             \altaffilmark{6},
        S.~Frey                   \altaffilmark{8},
        L.~I.~Gurvits             \altaffilmark{9},
        M.~L.~Lister              \altaffilmark{10},
        D.~W.~Murphy              \altaffilmark{11},
        Z.~Paragi                 \altaffilmark{9},
        B.~G.~Piner               \altaffilmark{12},
        Z.-Q.~Shen                   \altaffilmark{13},
	A.~R.~Taylor              \altaffilmark{5},
        S.~J.~Tingay              \altaffilmark{3},
        Y.~Asaki                  \altaffilmark{6}, 
        D.~Moffett                 \altaffilmark{14},
        Y.~Murata                 \altaffilmark{6}
}

\altaffiltext{1}{Australia Telescope National Facility,
                 Commonwealth Scientific and Industrial Research Organization,
                 P. O. Box 76, Epping NSW 2122, Australia}

\altaffiltext{2}{National Radio Astronomy Observatory, 
                 P.O. Box 0, Socorro, NM 87801, USA}

\altaffiltext{3}{Centre for Astrophysics and Supercomputing, 
                Swinburne University of Technology, P.O. Box 218, 
                Hawthorn, Victoria 3122, Australia}

\altaffiltext{4}{National Radio Astronomy Observatory, 520 Edgemont Road,
                 Charlottesville, VA 22903, USA}

\altaffiltext{5}{Physics and Astronomy Department, University of Calgary,
                 2500 University Dr. NW,
                 Calgary, Alberta, Canada, T2N 1N4}

\altaffiltext{6}{Institute of Space and Astronautical Science, 
                 Japan Aerospace Exploration Agency,
                 3-1-1 Yoshinodai,
                 Sagamihara, Kanagawa 229-8510, Japan}

\altaffiltext{7} {Dominion Radio Astrophysical Observatory,
                   P.O. Box 248, White Lake Road,
                   Penticton, BC, Canada
                   V2A 6K3}

\altaffiltext{8}{F\"{O}MI Satellite Geodetic Observatory, P.O. Box 585,
                 H-1592, Budapest, Hungary}

\altaffiltext{9}{Joint Institute for VLBI in Europe, P.O. Box 2,
                 7990 AA, Dwingeloo, The Netherlands} 

\altaffiltext{10}{Physics Department, Purdue University,
  525 Northwestern Ave, W. Lafayette IN, 47907-2036}

\altaffiltext{11}{Jet Propulsion Laboratory, 4800 Oak Grove Drive,
                 Pasadena, CA 91109, USA}

\altaffiltext{12}{Department of Physics \& Astronomy, Whittier College, 
       13406 E. Philadelphia St., Whittier CA 90608, USA}

\altaffiltext{13}{Shanghai Astronomical Observatory,
  80 Nandan Rd., Shanghai 200030, P.R.China}

\altaffiltext{14}{Physics Department, Furman University,
       3300 Poinsett Highway, Greenville, SC 29613, USA}

\begin{abstract} 
The VSOP mission is a Japanese-led project to study radio sources with
sub-milliarcsecond angular resolution using an orbiting 8-m telescope,
HALCA and global arrays of Earth-based telescopes.  Approximately 25\% of the
observing time has been devoted to a survey of compact AGN at 5 GHz
which are stronger than 1 Jy --- the VSOP AGN Survey.  This paper, the
second in a series, describes the data calibration, source detection,
self-calibration, imaging and modeling, and gives examples illustrating
the problems specific
to space VLBI. The VSOP Survey web-site which
contains all results and calibrated data is described.

\end{abstract}


\section{Introduction}

On 1997 February 12, the Institute of Space and Astronautical Science
(ISAS) launched the HALCA satellite carrying an 8\,m radio telescope
dedicated specifically to Very Long Baseline Interferometry (VLBI).
With an apogee height of 21400\,km, radio sources are able to be
imaged with angular resolution three times greater than with
Earth-based arrays at the same frequency \citep{hir98}.  About 25\% of
the observing time to date has been dedicated to the VSOP AGN Survey of
$\approx 400$ flat-spectrum AGN which are stronger than 1\,Jy at 5
GHz.  Observations from the VLBA Pre-Launch Survey \citep[hereafter VLBApls,][]{fom00b}
revealed that 294 of these sources demonstrated compact structures
suitable for observations with HALCA, and these were included in the
VSOP Source Sample (VSS). (This number was initially reported as 289
[\cite{hir00b}] but increased to 294 when it was found that the use of
low accuracy positions had initially resulted in 5 other sources not
being detected [\cite{2002aprm.conf..375E}].) The compilation and
general description of the VSOP AGN Survey is given by \cite{hir00b}
(Paper~I) and \cite{fom00a}.  The major goal of the Survey is to
determine statistical properties of the sub-milliarcsecond structure
of the brightest extragalactic radio sources at 5 GHz, and to compare
these structures with other properties of the sources.  Combined with
ground observations at many radio frequencies (single-dish and VLBI)
and at higher energies, the Survey will provide an invaluable source
list for detailed ground-based studies, as well as list of sources for
future space VLBI missions.

In this paper, the second in the VSOP Survey series, we describe the
data calibration and imaging procedures adopted for the VSOP Survey
Program.  These procedures are sufficiently different from more
conventional VLBI data reduction because of the relatively poor phase
stability and low signal-to-noise inherent in space VLBI\@.  Paper~III
\citep{sco04} presents results for the first 102 sources and Paper~IV
\citep{hor04} contains a statistical analysis using the visibility
data for sources with $\delta > -44^{\circ}$.

In \S2 we briefly review the correlation of VSOP data.  In \S3 we
discuss the calibration procedure, and in \S4 we outline the self-calibration, imaging and modeling of the sources.  Finally, in \S5 we
describe the VSOP web site and its access.

\section{Correlation of VSOP Data}

The VLBI Space Observatory Programme (VSOP) was described in detail by
\cite{hir00a, hir00b}.  For the Survey, the VLBI wavefront data are
recorded in the standard HALCA continuum mode at each participating
ground telescope, and HALCA data are similarly recorded by one or more
of the five tracking stations (see Paper~I).  The delays in the
downlink from the spacecraft to each tracking station were also
monitored at the tracking stations \citep{hir00a}.  Four recording
formats have been used in VSOP observations: VLBA \citep{nap94}, MkIV
\citep{whit}, S2 \citep{car99} and VSOP \citep{shi98}.  For many
Survey observations, a mixture of recording formats are used at
the tracking stations and ground telescopes; in these
cases, special-purpose devices located at the VSOP correlator in
Mitaka, Japan are used to translate the data to a common format, which
is then supplied to the appropriate correlator.  The majority of the
Survey experiments were correlated with the S2 Correlator (DRAO,
Penticton, BC, Canada) until 2002 August.  The VSOP correlator (NAOJ,
Mitaka, Japan) has been used for some of the observations which
included the ground telescopes at Usuda and Kashima, and has been used for all
observations after 2002 August.  The VLBA correlator (NRAO, Socorro,
NM, USA) was used until early 2002 for many of the General
Observing Time (GOT) experiments in VLBA and Mk4 formats, from which a
sub-set of the data was extracted for use in the Survey
\citep{hir00b}.  Data are exported from each correlator in a
format appropriate for initial reduction in the NRAO AIPS package
\citep{gre88}.

The correlator output consists of typically 128 frequency channels in
each of two 16~MHz bands. The VLBA and Penticton correlators produced
data at time-samplings of 0.5 seconds and 2.0 seconds for space-ground
and ground-only baselines respectively while the Mitaka correlator
produces data at a 1 second sampling on all baselines.  Thus the data have
sufficient resolution to search for fringes within a window
spanning a residual delay of $\pm 4 \mu$sec and a residual phase-rate
of 1~Hz.  This corresponds to a position error of 500~m and velocity
error of 3 cm/s for the HALCA satellite, significantly larger than the
nominal errors of the orbit determination \citep{hir00a}.

The translation integrity and the relative amplitude scaling of the
three correlators were checked using the results of several three-hour
experiments in which a strong source was observed using three ground
stations, two of which could record data simultaneously in two
formats.  The data were processed through all three correlators (with
format translations made as necessary), and the results were compared.
First, it was found that the translation process did not change the
correlated amplitude by more than 2\%, except when there were clear
indications of recording problems.  Second, the comparison of the
visibility amplitudes for the same experiment processed by each
correlator established the relative correlator amplitude scale factor
to an accuracy of 3\% (G.~A. Moellenbrock et al. 2002, private communication).

\section{Data Calibration and Detection}

The reduction of VSOP Survey observations is being undertaken by a
global effort of astronomers with an interest in high-resolution
imaging.  Therefore, a set of reduction procedures has been
developed to ensure, as much as possible, that the Survey results are
internally consistent.

The reduction procedure consists of two main parts.  The first part,
covered in this section, consists of initial calibration and
fringe-fitting, and is performed using the NRAO AIPS package.  The
second part, covered in \S4 consists of self-calibration, imaging and
model-fitting, and is performed using the Caltech Difmap package
\citep{she97}.  The following sections describe these steps in detail
with AIPS tasks and Difmap commands indicated by text in the {\sc
small caps} style.

\subsection{Preliminaries and {\em a priori} Calibrations.}  

The correlator distribution data for each experiment is imported into
NRAO AIPS using the task {\sc fitld}.  The datasets are sorted,
indexed, and documented using standard AIPS tasks ({\sc msort, indxr,
listr, prtan, dtsum}).  Except for datasets correlated in Penticton,
it is necessary to run {\sc accor} to remove fringe normalization
errors arising from potentially non-optimal sample populations among
the four 2-bit voltage levels recorded at each
telescope.  

For {\em a priori} amplitude calibration, system temperature and gain
information supplied by each telescope are imported into the AIPS
database using {\sc antab}.  Then, {\sc apcal} is used to form the
$\sqrt{\mbox{SEFD}}$ calibration factors required to scale each antenna's
gain\footnote{The {\em System Equivalent Flux Density (SEFD)} is the
ratio of the system temperature (K) and telescope gain (K/Jy).  It
concisely describes the sensitivity of a radio telescope and the
geometric mean of SEFD for two telescopes and provides the proper
scaling factor to convert normalized correlation coefficients to
Janskys.}.  For HALCA, the nominal 5~GHz system temperature is
$\sim90$ K and stable within an observation to $\sim 5$\%.  Its SEFD was
monitored early in the mission using total power observations of
Cas~A, Cyg~A or Tau~A and found to be relatively constant.  The 5~GHz
gain is 0.0062 K/Jy and this yields a HALCA SEFD of $\sim14,500$ Jy, which
is more than an order of magnitude larger than most ground telescopes.
The {\em a priori} amplitude calibration value and reliability from
the ground telescopes varies considerably, and can be in error by 30\%
for telescopes which are only occasionally used for VLBI.  

As VSOP observations are made with global arrays of ground telescopes,
it is not uncommon for some telescopes to be observing at frequencies
in the 5~GHz band offset from the their standard frequencies, i.e.,
the frequencies at which the nominal gain is measured and
monitored. HALCA's 5~GHz system noise temperature varies by almost
15\% across the 4700--5000 MHz band \citep{kob00} and so Survey
observations are generally scheduled at the frequencies where (the
least sensitive telescope) HALCA's performance was best. Use of
nominal gain and nominal system temperature values, or even measured
system temperature values if these were measured at the standard
frequencies rather than the actual observing frequencies, also
contributed to the overall uncertainties in gain calibration of VSOP
Survey observations.  Further amplitude corrections are discussed in
the next sections, and Paper III describes a more accurate post-facto
determination of the amplitude calibration of the Survey sources as a
whole.

\subsection{Fringe-fitting} 

Fringe-fitting, the process by which the correlated signals are
detected, is the most important part of the Survey reductions.  Unlike
most ground-only VLBI, fringe detection for HALCA observations is
difficult due to the limited sensitivity of the orbiting telescope,
the generally lower correlated flux densities on long baselines and
the larger uncertainty in the spacecraft's location and clock.  (The
``spacecraft clock'' is the hydrogen maser at the tracking station in
use at the time, however the corrections required to correct for the
downlink of the data introduced uncertainties in addition to those
encountered for ground radio telescopes [Hirabayashi et al. 2000a].)
This combination of conditions requires fringe searches for weak
signals over large ranges of delay and fringe rate, hence the need for
high time and frequency sampling.  The small number of ground
telescopes typical of Survey observations limits the sensitivity for
global fringe-fitting as well \citep{cot95}.  It is therefore
important to limit the range of the search in delay and fringe rate as
much as possible in order to keep the fringe searching efficient, and
to avoid false detections.  For many Survey datasets, delay and fringe
rate solutions are available from fringe searches performed for
data quality analysis at the correlators.  Application of these delay and
phase-rate offsets using {\sc clcor} before more detailed fringe
searching allows for significant data averaging and smaller fringe-search
windows.  For datasets with strong fringes for only a portion of the
observation, the resulting narrower search should, in principle, allow
detection of the weaker fringes.  In practice, however, these gains
have been modest.

\begin{figure}
%\special{psfile=f1.eps hoffset=15 voffset=-500 hscale=70 vscale=70 angle=0}
%\vspace{100mm}
\plotone{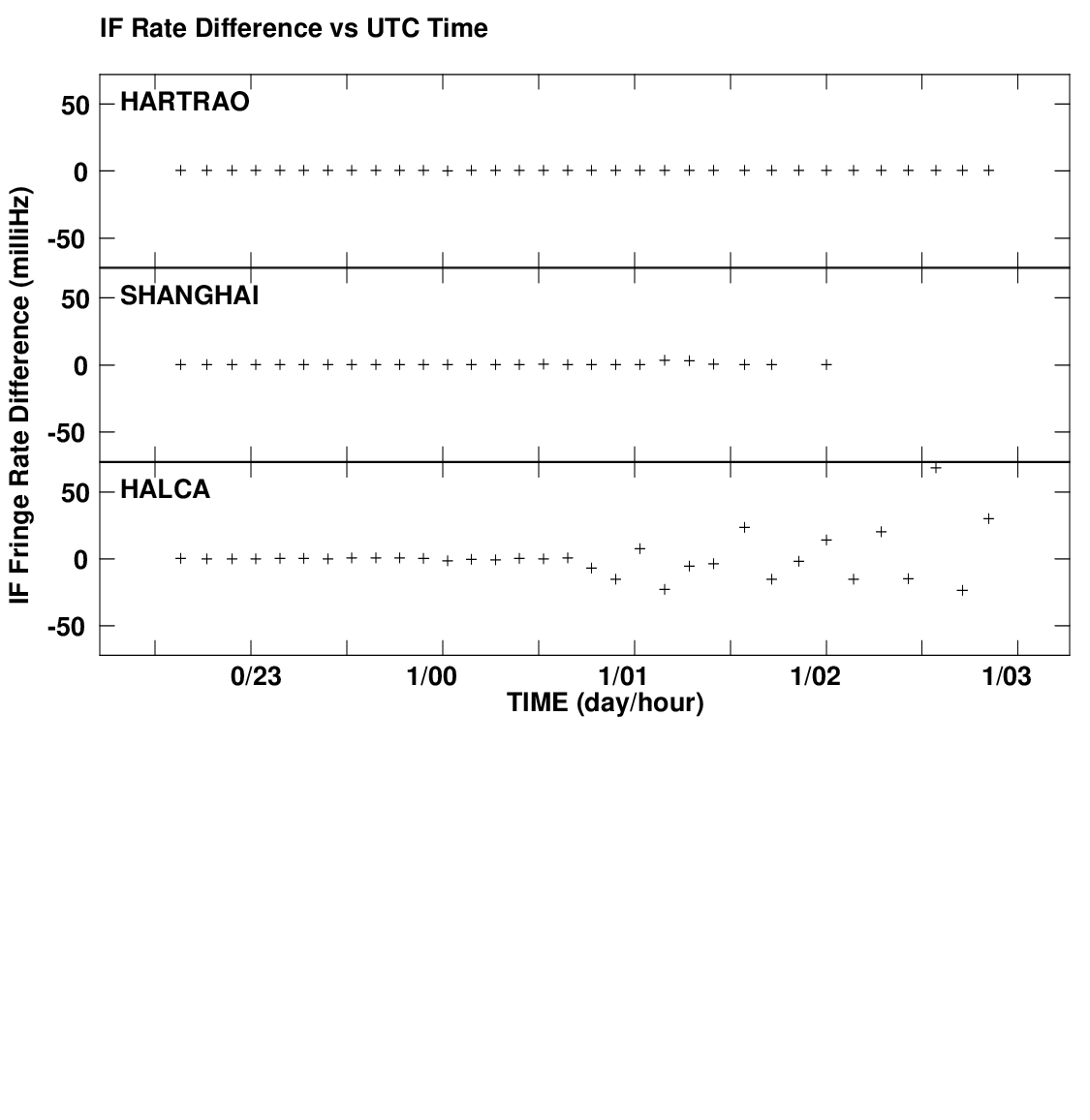}
\caption{An example of the difference in fringe rate solutions between
the two bands for a typical Survey experiment.  Fringes to HALCA are
detected until nearly 1/01 at which time the differences become
randomly distributed on a scale exceeding the fringe rate resolution
($\sim$3\,mHz FWHM). In this case the fringe rate search window was
restricted which is why the rate differences are constrained after
fringes are lost.}
\label{ratediff}
\end{figure}

For most observations, the AIPS task {\sc fring} is used for
fringe-fitting.  Solution intervals of 
up to 10 minutes (approaching the coherence time) are attempted to
maximize the Signal-to-Noise Ratio (SNR).  For strong sources, solution intervals as short as 2
minutes can be used as long as the SNR is greater than about 5.
Detections are best gauged by consistency in the delay and fringe rate
solutions between the two independent frequency channels (see Figure
\ref{ratediff}).  For the weakest sources (few or no detections in the
correlator's data quality search), the AIPS task {\sc kring} was used
since it allows larger searches and longer integrations than {\sc
fring} for the same computer resources.  Most of the editing of the
data was obvious from the loss of detection during the fringing
process, and from the telescope logs.  

After an adequate fringe-fitting solution is obtained, the combined
calibration was applied using {\sc split} or {\sc splat}, which also averages the
corrected data in frequency within each 16 MHz band, and to a common
2-second integration time.  The data were then stored using {\sc
fittp} or {\sc fitab} for subsequent processing.

\section{Determining the Source Structure}

In their analysis of VSOP observations of a complete sample of
northern sources with very good $(u,v)$ coverage on ground-only
baselines, Lister et al. (2001) found that the dynamic range of a VSOP
image is limited by poor sampling of the $(u,v)$ plane on ground-space
baselines. They found that the true dynamic range is between 30:1 and
100:1 depending on source complexity. In the case of VSOP Survey data
this problem is amplified by the smaller number of ground radio
telescopes and great care must be taken in interpreting the
data. Although every effort was made to include long ground baselines
in the scheduled array, this was not always achievable in practice. In
general, VSOP Survey data provides a general idea of structure such as
core sizes and intensities and basic jet properties such as position
angle and location with respect to the core.

It is therefore important for the data analyst to keep these
limitations in mind when working with this sparsely sampled
$(u,v)$-data with low SNR.  Each stage of the data reduction must be
checked in order to obtain a source structure which is consistent with
the limitations in the data as well as incorporating {\em a priori}
information about the source structures, either from the ground-only
baselines or from other ground-based VLBI observations of the source.
The Difmap software package was chosen for this part of the processing
since it provides a good interface for viewing data, as well as
visibility-plane model-fitting and image deconvolution facilities.

\subsection {First-Pass Editing and Checking}

The data from the AIPS calibration (2~s sampling in two single-channel
frequency bands) are read into Difmap and averaged to a 30~s grid. The
weights are calculated as the reciprocal of the data variance, which
is proportional to the inverse square of the RMS. The data are then
phase self-calibrated with a point source model on a 30~s timescale to
determine the telescope-based phase as a function of time.  Further
data editing is based on several criteria: (1) Obvious outliers in a
plot of amplitude versus projected $(u,v)$-distance are removed
using {\sc radplot}; (2) Periods of low visibility amplitude for any
antenna are found using {\sc vplot}; (3) Periods of very poor
phase stability (indicating that the source was not detected during
this period) can be seen using {\sc corplot}.

Although the {\em a priori} gain calibration for each telescope is made in
AIPS using the nominal gain and system temperatures, large residual,
gain errors of up to $30$\% often persist.  For this reason,
observations of additional compact sources by the ground telescopes
are scheduled (typically during gaps in HALCA tracking) and used to
better constrain the gain values for each telescope.  These
calibrators have known structures from the VLBApls
catalog \citep{fom00b} and their flux densities 
monitored from observing programs at the University of
Michigan\footnote{http://www.astro.lsa.umich.edu/obs/radiotel/umrao.html}
and at the Australia Telescope Compact Array
\citep{tin03}.

\subsection{Self-Calibration and Imaging Iterations}

Since most Survey experiments have limited $(u,v)$-coverage, imaging
and/or model-fitting requires the introduction of constraints to the
size and complexity of the radio emission in order to obtain accurate
deconvolution and self-calibration.  The first step is to make a
relatively low resolution image without the HALCA data.  These images
provide the best sensitivity to extended structure and help identify
regions in the field where the most compact structure was likely to be
located.  The VLBApls catalog image, made from observations in 1996
\citep{fom00b}, as well as other pre-existing images of the sources
(including ground observations at 15 GHz which had similar resolution
of the VSOP Survey observations [Kellermann et al. 1998; Gurvits et
al. 2004, in preparation]) are also useful in determining the
constraints needed to image the VSOP Survey data.  For about 5\% of
the VSOP Survey observations, it is clear that the visibility
amplitude on the shortest projected baselines is much lower than that
known from pre-existing VLBI observations, even considering possible
variability of sources. This indicates that a problem has occurred at
one or more antennas during the observation or during the tape copying
process (if it was required) or during correlation. If the problem can
not be rectified the observations are considered corrupted, and placed
back into the VSOP Survey observing schedule for another
observation. In cases such as these the data are processed, often as a
ground-only observation, as they may still provide useful information.

\begin{figure}
%\special{psfile=f2.ps hoffset=0 voffset=30 hscale=35 vscale=35 angle=270}
%\special{psfile=f3.ps hoffset=250 voffset=30 hscale=35 vscale=35 angle=270}
%\vspace{60mm}
\includegraphics[angle=270,scale=0.33]{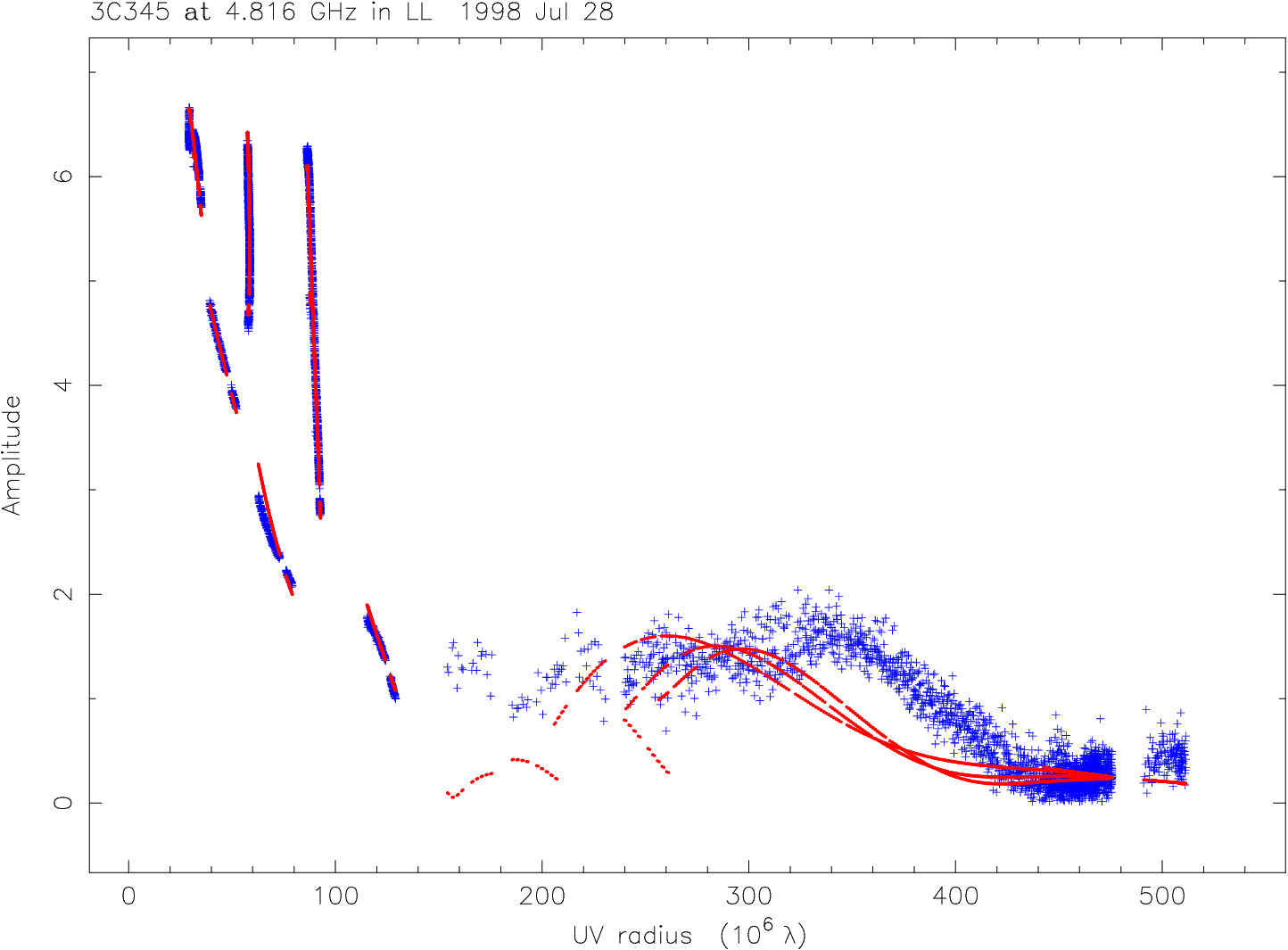}
\includegraphics[angle=270,scale=0.33]{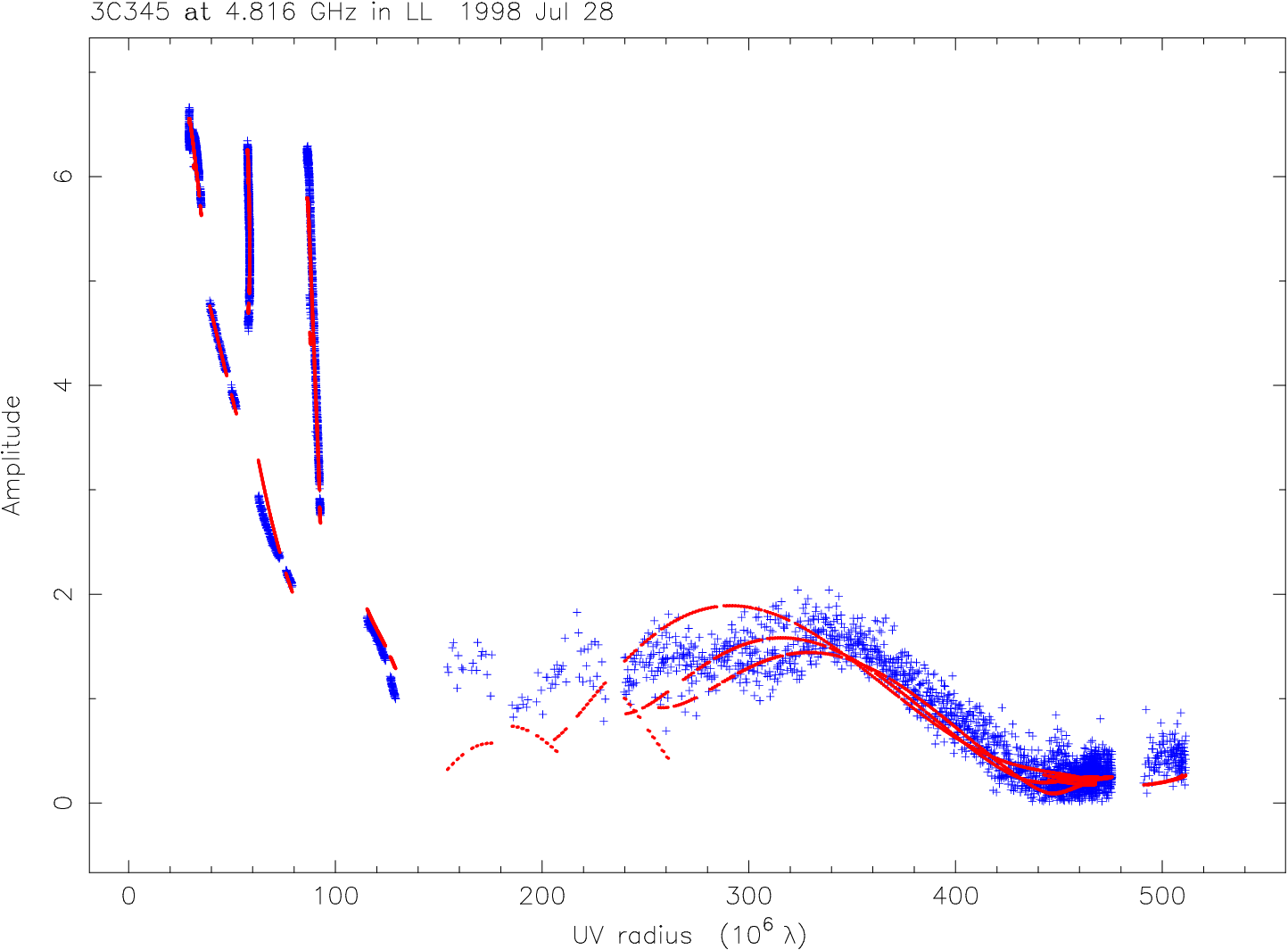}
\caption{Model fits to the visibility data from a VSOP Survey
observation of 3C345. Visibility amplitudes (+ symbols) and model
visibilities (solid points) are plotted as a function of $(u,v)$
radius. {\bf Left:} the result of a fit using the standard
$1/{\sigma^2}$ weighting and {\bf right:} a model fit with the weight
on HALCA data increased by a factor of 25.}

\label{fig:modfit}
\end{figure}

The next imaging step includes all of the data to obtain an
approximate image.  For most sources, the $(u,v)$-coverage is
sufficient to use CLEAN, followed by a phase-only self-calibration to
improve the phase calibration.  Several phase self-calibration
iterations are generally made for each source.  For sources with
extremely poor $(u,v)$-coverage, model-fitting the data with one
or two Gaussian-shaped components is used instead of the CLEAN
deconvolution.  In some cases a hybrid approach, using CLEAN
components for the extended emission and models for the small-diameter
components, is used.  The use of various data weightings to emphasize
or de-emphasize the longer VSOP baselines depends upon the strength
and size of the source and the number of visibilities on ground-only
baselines compared to ground-space baselines. To obtain images that
best reveal the $\sim 0.1$~mas scale a weighting scheme is used
for which the
space-ground baselines contribute about 50\% of the effective data.
As an illustration of the importance of
increasing the data weights on space baselines, we present the results
of fitting a simple model to the VSOP Survey observations of 3C345 on
1998 July 28 (Figure~\ref{fig:modfit}). The {\sc modelfit} procedure
in Difmap applies a weight of $1/{\sigma^2}$ to each visibility
point. When the weights are calculated in this way the sensitive
ground-only visibilities dominate the fit and the ground-space
baselines have little influence. However, if the HALCA data are
upweighted, in this case by a factor of 25 so that ground-space and
ground-only visibilities have roughly equal weighting, the fit
improves significantly.

For most Survey experiments, amplitude self-calibration is not used due
to lack of closure constraints and/or limited sensitivity of the
space-baselines.  In the cases where the data from four or more
telescopes are sufficiently strong, amplitude self-calibration using
{\sc gscale} provides a scale factor for each telescope over the
entire observation.  In some cases, amplitude self-calibration over a
time-scale of one hour is possible.

\subsection{Image Representation}
\label{sec:modfit}

A satisfactory image is generally obtained after three or four phase
self-calibration loops and perhaps one amplitude calibration. Such a
quick convergence is due to the limitations on the achievable image
fidelity characterised by a sparse $(u,v)$ coverage, and a lack of
short spacings in particular. The latter typically contain most of the
information on complex, extended structures.  For most sources two
representations of the source structure are available: the CLEAN
image, and a model-fit image.  For small-diameter components with poor
$(u,v)$-coverage, the model representation is more reliable than the
CLEANed image.  In some cases, the CLEAN components more accurately
reproduce the extended emission, while the Gaussian model component
describes the small-diameter components more accurately.  Both
representations of the source structure should be in relatively good
agreement, and satisfactorily fit the observed $(u,v)$-data.

The model representation of the source structure isolates important
parameters of the components which are necessary to determine angular
sizes and brightness temperatures.  Even for images in which the CLEAN
algorithm was used to determine the source structure, the models are
chosen carefully, starting simple and moving to more complicated
structures.  The goal is to fit the observed visibilities within the
uncertainties using the smallest number of model parameters, and to
duplicate the structure found by cleaning.  The following guidelines
are used in choosing model components:

\begin{enumerate}

\item The number of components is kept to a minimum.

\item Simple components are preferred to complex ones, {\em i.e.} a
point model is better than a circular Gaussian model which is better than an
elliptical Gaussian component.

\item If an elliptical component becomes linear during model fitting it
was generally an indication that the sampling of the $(u,v)$-plane
was poorly constrained in the direction perpendicular the the
component's major axis. In these cases, a circular Gaussian component
is favored.

\item In general, additional or more complex components are used only
if the data or the image require it.

\end{enumerate}

The calibrated data, models and cleaned image are then saved in the
NRAO AIPS UVFITS format using Difmap's {\sc save} command.
These data can be read back into Difmap for further processing,
imaging and modeling, and are available through the VSOP Survey Data
Base Web site.

\section{VSOP Survey Data Base}
Once a data analyst has completed the reduction of an experiment, the
calibrated data, reduction notes and supporting files (from both AIPS
and Difmap) are uploaded to a computer at ISAS. Information from the
uploaded files is entered into a database and published on the VSOP
Survey web page (http://www.vsop.isas.jaxa.jp/survey).  Displays of
the final calibrated visibilities, images (CLEAN and model-fit), and
the model fit parameters are available.  Documentation of the data
processing, from the initial calibration, to the fringe-fitting, to
the imaging and modeling, are given for each experiment.  The
calibrated data are available from the web site and the original
post-correlation data can be obtained upon request, but this dataset
is much larger (of the order of 100~Mb compared to $\sim 100$~kb).

\section{Conclusion}
This paper, the second in the VSOP 5 GHz AGN Survey series, has
described the data processing used to construct the images and determine models
from the VSOP Survey Program.  Because of the uniqueness of the
Space VLBI Survey data, we have described many of the procedures in some detail
since they are different from normal VLBI reduction practices.
Enhancement of the data procedures and the development of new
algorithms (especially for detecting weak sources) are needed for
further space VLBI missions.

\acknowledgements
The National Radio Astronomy Observatory is a facility of the National
Science Foundation operated under cooperative agreement by Associated
Universities, Inc.  This research has made use of data from the
University of Michigan Radio Astronomy Observatory, which is supported
by funds from the University of Michigan.  J.E.J.L., G.A.M., and
R.G.D. each acknowledge the support of Japan Society for the Promotion
of Science fellowships. W.K.S. and A.R.T. wish to acknowledge support from the Canadian Space
  Agency.

\clearpage

\newpage


\begin{thebibliography}{}

\bibitem[\protect\citeauthoryear{Carlson et al.}{1999}]{car99}
Carlson, B. R., Dewdney, P. E., Burgess, T. A., Casorso, R. V.,
Petrachenko, W. T., \& Cannon, W. H.
1999,
\pasp,
111, 1025

\bibitem[\protect\citeauthoryear{Cotton}{1995}]{cot95} Cotton,
W.~D. 1995, in ASP Conf. Ser. 82, Very Long Baseline Interferometry
and the VLBA, eds. J.~A.~Zensus, P.~J.~Diamond and P.~J.~Napier (San
Francisco: ASP), 189

\bibitem[Edwards et al.(2002)]{2002aprm.conf..375E} Edwards, P.~G., 
Hirabayashi, H., Fomalont, E.~B., Gurvits, L.~I., Horiuchi, S., Lovell, 
J.~E.~J., Moellenbrock, G.~A., \& Scott, W.~K.\ 2002, 8th Asian-Pacific 
Regional Meeting, Volume II, 375 

\bibitem[\protect\citeauthoryear{Fomalont et al.}{2000a}]{fom00a}
Fomalont, E., et al.
2000a,
in Proceedings of the VSOP Symposium, January 2000a,
Astrophysical Phenomena Revealed by Space VLBI,
eds. H. Hirabayashi, P. G. Edwards, \& D. W. Murphy
(Sagamihara: The Institute of Space and Astronautical Science),
167

\bibitem[\protect\citeauthoryear{Fomalont et al.}{2000b}]{fom00b} 
Fomalont, E. B., Frey, S., Paragi, Z., Gurvits, L. I., Scott, W. K.,
Taylor, A. R., Edwards, P. G., \& Hirabayashi, H.
2000b,
\apjs,
131, 95

\bibitem[\protect\citeauthoryear{Greisen}{1988}]{gre88}
Greisen, E. W.
1988,
in Acquisition, Processing and Archiving of Astronomical Images, 
eds. G. Longo \& G. Sedmak
(Napoli: Osservatorio Astronomico di Capodimonte), 
125

\bibitem[\protect\citeauthoryear{Horiuchi et al.}{2004}]{hor04}
Horiuchi, S., et al., 2004, \apj, submitted.

\bibitem[\protect\citeauthoryear{Hirabayashi et al.}{1998}]{hir98}
Hirabayashi, H., et al. 
1998, 
Science, 
281, 1825

\bibitem[\protect\citeauthoryear{Hirabayashi et al.}{2000a}]{hir00a}
Hirabayashi, H., et al.  2000a, \pasj, 52, 6, 955

\bibitem[\protect\citeauthoryear{Hirabayashi et al.}{2000b}]{hir00b}
Hirabayashi, H., et al.  2000b, \pasj, 52, 6, 997

%\bibitem[\protect\citeauthoryear{Hirabayashi}{2000c}]{hir00c}
%Hirabayashi, H.
%2000c,
%Adv. Space Res.,
%26, 751

\bibitem[Kellermann, Vermeulen, Zensus, \& 
Cohen(1998)]{1998AJ....115.1295K} Kellermann, K.~I., Vermeulen, R.~C., 
Zensus, J.~A., \& Cohen, M.~H.\ 1998, \aj, 115, 1295 

\bibitem[\protect\citeauthoryear{Kobayashi et al.}{2000}]{kob00}
Kobayashi, H., et al.  
2000,
\pasj,
52, 6, 967


\bibitem[Lister et al.(2001)]{2001ApJ...554..948L} Lister, M.~L., Tingay, 
S.~J., Murphy, D.~W., Piner, B.~G., Jones, D.~L., \& Preston, R.~A.\ 2001, 
\apj, 554, 948 

\bibitem[\protect\citeauthoryear{Napier et al.}{1994}]{nap94}
Napier, P. J., Bagri, D. S., Clark, B. G.,
Rogers, A. E. E., Romney, J. D., Thompson, A. R., \&
Walker, R. C.
1994,
IEEE Proc. 82,
658

\bibitem[\protect\citeauthoryear{Scott et al.}{2004}]{sco04}
Scott, W. K., et al., 2004, \apjs, accepted.

\bibitem[\protect\citeauthoryear{Shepherd}{1997}]{she97}
Shepherd, M. C.
1997,
in ASP Conf. Ser. 125,
Astronomical Data Analysis Software and Systems VI,
eds. G. Hunt, \& H.E. Payne
(San Francisco: ASP),
77

\bibitem[\protect\citeauthoryear{Shibata et al.}{1998}]{shi98}
Shibata, K. M., Kameno, S., Inoue, M., \& Kobayashi, M. 1998,
in ASP Conf. Ser. 144, 
Radio Emission from Galactic and Extragalactic Compact Radio Sources,
eds. J. A. Zensus, G. B. Taylor, \& J. M. Wrobel
(San Francisco: ASP),
397

\bibitem[\protect\citeauthoryear{Tingay et al.}{2003}]{tin03}
Tingay, S. J., Jauncey, D. L., King, E. A., Tzioumis, A. K., 
Lovell, J. E. J., \&  Edwards,  P. G.
2003, \pasj, 55, 351


\bibitem[\protect\citeauthoryear{Whitney}{1999}]{whit}
Whitney, A. R. 1999, New Astron. Rev., 43, 527


\end{thebibliography}
\end{document}